\documentclass{vgtc}                        

\ifpdf
  \pdfoutput=1\relax                   
  \usepackage{graphicx}                
  \DeclareGraphicsExtensions{.pdf,.png,.jpg,.jpeg} 
\else
  \usepackage{graphicx}                
  \DeclareGraphicsExtensions{.eps}     
\fi%

\graphicspath{{figures/}{pictures/}{images/}{./}} 

\usepackage{microtype}                 
\PassOptionsToPackage{warn}{textcomp}  
\usepackage{textcomp}                  
\usepackage{mathptmx}                  
\usepackage{times}                     
\usepackage{cite}                      
\usepackage{tabu}                      
\usepackage{booktabs}                  
\usepackage{enumitem}

\onlineid{1122}

\vgtccategory{Application}

\vgtcinsertpkg

\usepackage{xspace}

\newcommand{\myparagraph}[1]{\vspace{1mm} \noindent \textbf{#1}}

\newcommand{\etal}{et al.\xspace}

\title{Segmentation Driven Peeling for Visual Analysis of Electronic Transitions}

\author{Mohit Sharma$^1$\thanks{Email: \{mohitsharma,vijayn\}@iisc.ac.in}%
\hspace{7pt} Talha Bin Masood$^2$%
\hspace{7pt} Signe S. Thygesen$^2$%
\hspace{7pt} Mathieu Linares$^{2}$%
\hspace{7pt} Ingrid Hotz$^2$%
\hspace{7pt} Vijay Natarajan$^1$\\ %
     \parbox{5in}{\small \centering $^1$Indian Institute of Science, Bangalore, India \\ $^2$Department of Science and Technology (ITN), Link\"oping University, Norrk\"oping, Sweden}}
\teaser{
  \centering
  \includegraphics[width=\linewidth]{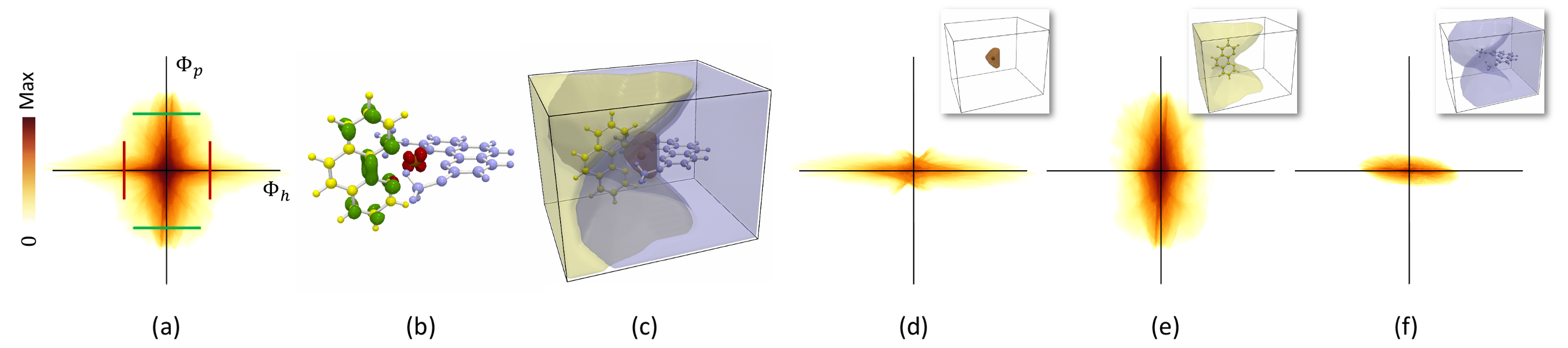}
  \caption{Studying electronic transitions in a copper complex Cu-PHE-PHEOME. (a)~Continuous scatter plot (CSP) of the hole and particle NTO ($\phi_h$, $\phi_p$) and query line segments (green and red). (b)~Fiber surfaces corresponding to the query line segments displayed on a ball-and-stick model of the molecule. (c)~A Voronoi diagram-based segmentation of the molecule into its subgroups. (d,e,f)~Peeled CSP corresponding to the subgroups (Cu, PHE, PHEOME; inset - Voronoi regions) help study their behavior.}
  \label{fig:teaser}
}

\abstract{Electronic transitions in molecules due to absorption or emission of light is a complex quantum mechanical process. Their study plays an important role in the design of novel materials. A common yet challenging task in the study is to determine the nature of those electronic transitions, i.e. which subgroups of the molecule are involved in the transition by donating or accepting electrons, followed by an investigation of the variation in the donor-acceptor behavior for different transitions or conformations of the molecules. In this paper, we present a novel approach towards the study of electronic transitions based on the visual analysis of a bivariate field, namely the electron density in the hole and particle Natural Transition Orbital (NTO). The visual analysis focuses on the continuous scatter plots (CSPs) of the bivariate field linked to their spatial domain. The method supports selections in the CSP visualized as fiber surfaces in the spatial domain, the grouping of atoms, and segmentation of the density fields to peel the CSP. This peeling operator is central to the visual analysis process and helps identify donors and acceptors. We study different molecular systems, identifying local excitation and charge transfer excitations to demonstrate the utility of the method.%
} 

\CCScatlist{
  \CCScatTwelve{Human-centered computing}{Visualization}{Visualization application domains}{Scientific visualization};
}

\nocopyrightspace

\begin{document}

\maketitle

\section{Introduction}
The study of electronic transitions in molecules due to absorption or emission of light, called molecular spectroscopy, is crucial for understanding their chemical and physical properties~\cite{Kim2019}. Electrons are distributed within a series of available orbitals in a molecule~\cite{Mulliken}. When a photon is absorbed by a molecule, electrons are excited from occupied orbitals to unoccupied orbitals resulting in a change in the electronic structure of the molecule. The process is reversed for emission of light. The orbital vacated by the electron is named as \emph{hole} and the orbital that gets filled by the electron is named as \emph{particle}.  The Natural Transition Orbital (NTO) is a compact representation of electronic excitations. It considers a linear combination of orbitals involved in a specific electronic transition to describe from where the electrons are excited  (hole NTO) and to where they are promoted (particle NTO)~\cite{Martin2003NTO}. 

Our objective is to develop a solution for a problem of interest to a theoretical chemist, namely the analysis of the nature of electronic transitions in molecules. A typical study of electronic transitions poses questions related to the transfer of charge within and between molecular subgroups such as: 
Which subgroup is a donor / acceptor for a particular excited state of the molecule? 
How do the donor / acceptor strengths (charge transfer) vary with different molecular conformations? 
How do donor / acceptor behavior vary within a family of complexes?
Can a given state be classified as presenting a local excitation or charge transfer character?

\begin{figure*}
  \centering
  \includegraphics[width=0.9\linewidth]{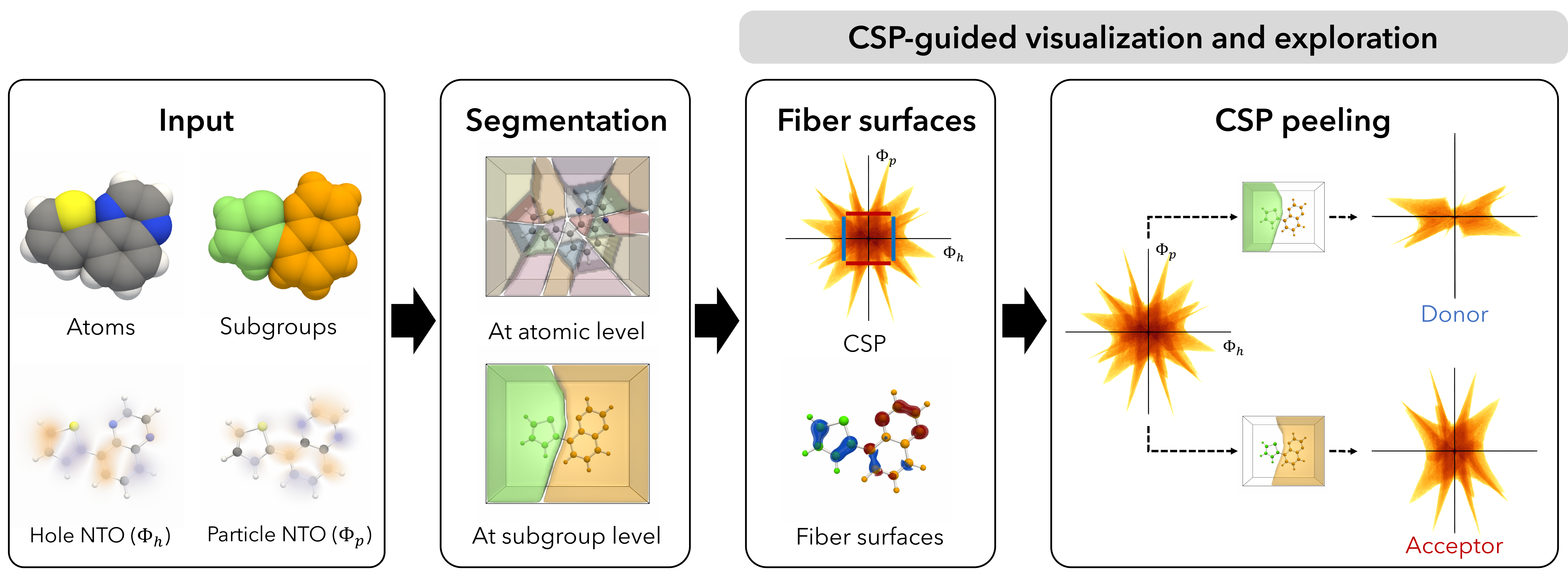}
\vspace{-0.4em}
\caption{Visual analysis workflow. Molecular structure, electron density fields, and subgroup descriptions are available as input. A weighted Voronoi segmentation identifies atomic regions. The segment corresponding to a molecular subgroup is computed as the union of atomic regions. The CSP of the bivariate field may be explored either using fiber surfaces or peeled on demand based on individual segments corresponding to the subgroups. Peeled CSPs are visually analyzed side-by-side and in comparison with the CSP of the bivariate field for subgroup classification and donor-acceptor strength comparisons.}
\vspace{-1.2em}
\label{fig:pipeline}
\end{figure*}

\myparagraph{Related work.}
The electron density fields are often studied individually via isosurfaces together with ball-and-stick model representations of the molecule and additional annotations to indicate charge transfer~\cite{Humphrey1996,Stone2011}. 
These techniques are useful for the study of  orbitals within a given state and rely on side-by-side comparisons to understand transitions~\cite{Haranczyk2008}.
The analysis of transfer of charge between the molecular subgroups and their classification as donor-acceptor is a key step towards the study of electronic transitions. 
A few studies employ quantitative approaches towards the study of charge transfer. 
Garcia~\etal~\cite{Garcia2010,Bahers2011}
propose a set of indexes based on point wise difference in charge density field to capture the amount of charge transfer and change in dipole moment, and later also interpreted as hole-electron distance~\cite{Guido2013,Huet2020}.
In recent work, Masood \etal~\cite{masood2021visual} present an automated method for quantifying charge distribution and transitions based on a spatial segmentation of the electronic density field. 
They formulate charge transfer as a constrained optimization problem by modeling the molecular subgroups as nodes in a graph and describe methods for visualizing the charge distribution and transfer at the atomic and subgroup level.

Generalization of discrete scatter plots to continuous scalar fields~\cite{Bachthaler2008CSP} and of isosurfaces to fiber surfaces~\cite{carr2015fiber} have led to new directions in bivariate and multivariate field visualization. Carr \etal~\cite{carr2015fiber} highlighted the applicability of fiber surface to electron density fields. Subsequent work focused on improving the correctness and efficiency of fiber surface computation~\cite{klacansky2016fast} and various applications~\cite{tierny2016jacobi,Blecha2019Nuclear,Raith2019Tensor}. Tierny \etal~\cite{tierny2016jacobi}  segment the domain based on the Reeb space~\cite{edelsbrunner2008reeb} and use them to peel the continuous scatter plot (CSP) to reveal its connected structures.
Lehmann \etal~\cite{Lehmann2010} investigate discontinuities in CSPs to establish a relationship with the number of connected components in the spatial domain. 

We propose a new approach that analyzes the bivariate field consisting of the hole NTO and particle NTO to directly capture the nature of electronic transitions. 
We employ visualization techniques including CSP~\cite{Bachthaler2008CSP} and fiber surfaces~\cite{carr2015fiber} to explore the bivariate field both within the spatial domain and in the range space. The CSP may be queried using fiber surfaces to identify regions that have a strong donor or acceptor characteristic, see \autoref{fig:teaser}.
Further, we propose the use of smart peeling operations to explore the CSP. The peeling operator uses a segmentation of the spatial domain to generate CSPs corresponding to individual segments. This process can be imagined as peeling away different layers from the CSP to reveal contributions from different molecular subgroups of interest. The observation that the green and red fiber surfaces in \autoref{fig:teaser} identify individual molecular subgroups motivates the design of the segmentation driven peeling operator.

\vspace{-0.0em}
\myparagraph{Contributions.}
We introduce an approach towards the study of electronic transition based on visual analysis of a bivariate field, the hole and particle NTO. Key contributions of this paper include
\begin{enumerate}[noitemsep,topsep=0pt,leftmargin=0.4cm]
    \item A domain segmentation driven CSP peeling operator that enables exploration of the hole / particle NTO bivariate field.
    \item A visual analysis framework based on the peeling operator that supports the above-mentioned queries.
    \item Case studies on two molecular systems that demonstrate the utility and effectiveness of the framework.
\end{enumerate}
The first case study is composed of two rings namely thiophene and quinoxaline that are commonly used alone or combined to create conductive polymers for the field of organic electronics. Thiophene is well known for giving electrons while Quinoxaline accepts electrons.
For the second case study, we focus on a series of copper complexes used for luminescence applications. Those complexes are formed of a central copper atom surrounded by two ligands: one ligand is always the same in all complexes (Phenanthroline, PHE) while the other ligand varies: Phenanthroline (PHE), Dimethoxy Phenanthroline (PHEOME), and a diphosphine ligand (XANT). The character of the electronic transitions in those complexes will define their properties and it is important to be able to identify them easily to gain knowledge and help the design of novel candidates.
The method development, case study identification, and interpretations  were conducted in close collaboration with a theoretical chemist. 

\begin{figure*}[!t]
    \centering
    \begin{tabular}{c@{\hskip3pt}c@{\hskip1pt}c@{\hskip1pt}c@{\hskip1pt}c@{\hskip1pt}c@{\hskip5pt}c}
    & $0^{\circ}$ & $60^{\circ}$ & $90^{\circ}$ & $120^{\circ}$ & $180^{\circ}$ & Relaxed ($35.3^{\circ}$)
    \\
    \cmidrule(lr){2-6} \cmidrule(lr){7-7}
    \raisebox{0.2\height}{\rotatebox{90}{Geometry}}
    &
    \includegraphics[width=0.13\textwidth]{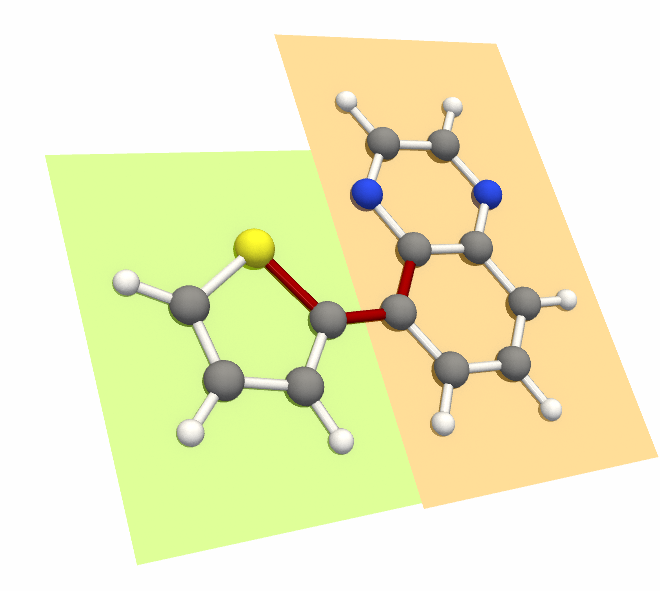}
    &
    \includegraphics[width=0.13\textwidth]{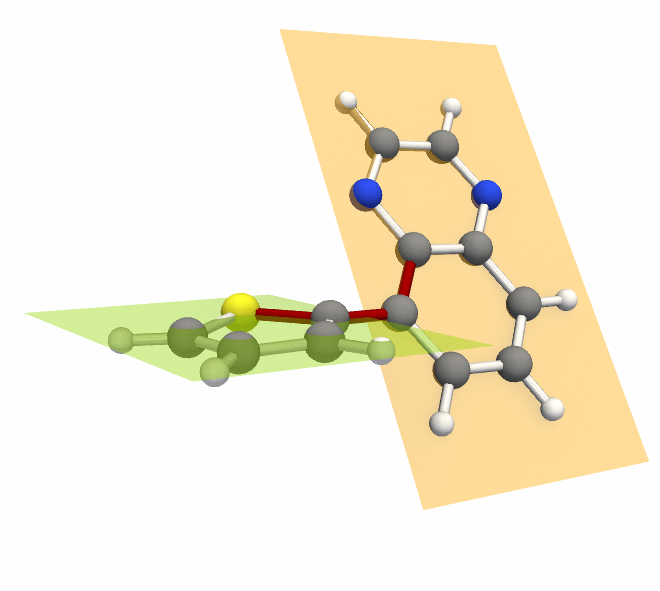}
    &
    \includegraphics[width=0.13\textwidth]{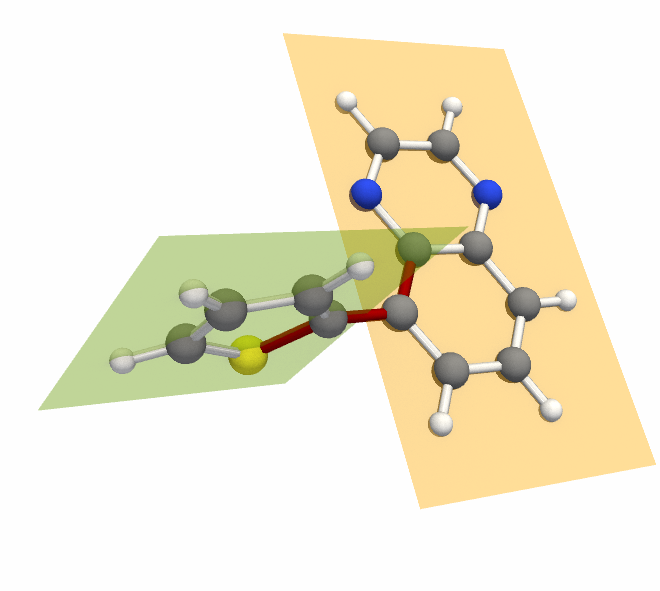}
    &
    \includegraphics[width=0.13\textwidth]{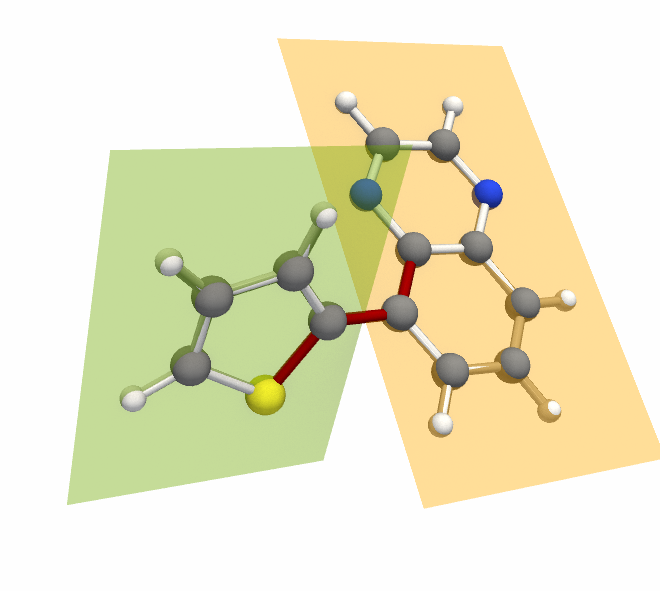}
    &
    \includegraphics[width=0.13\textwidth]{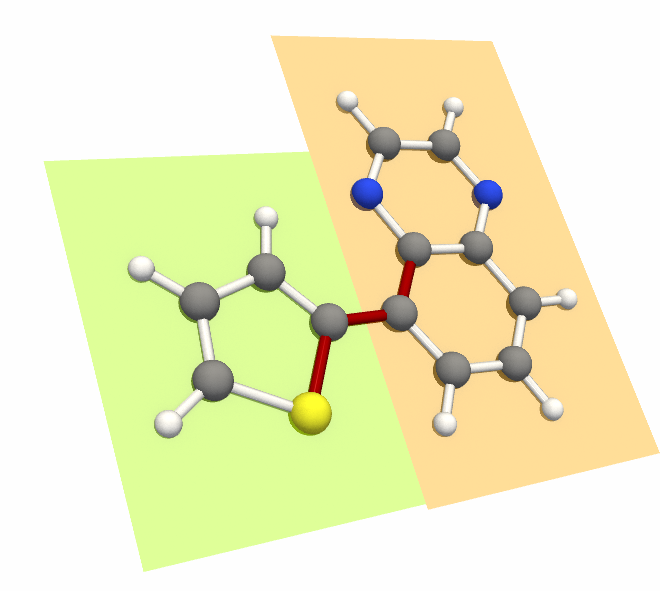}
    &
    \includegraphics[width=0.13\textwidth]{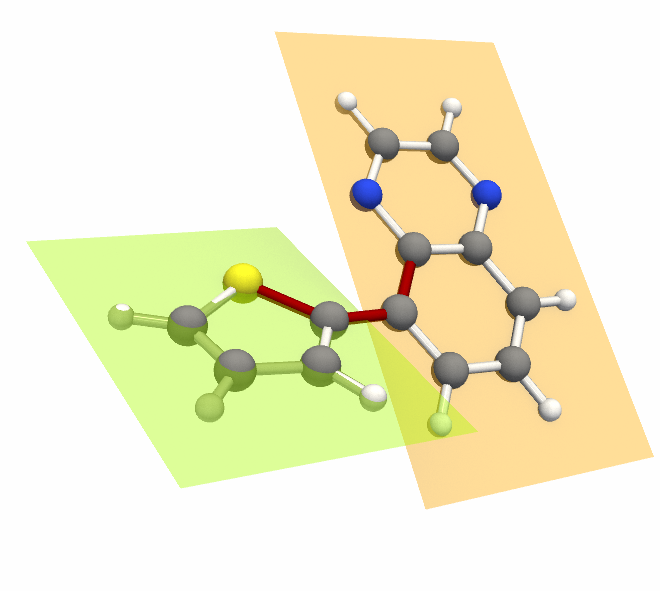}
    \\
    \raisebox{0.1\height}{\rotatebox{90}{Complete CSP}}
    &
    \includegraphics[width=0.13\textwidth]{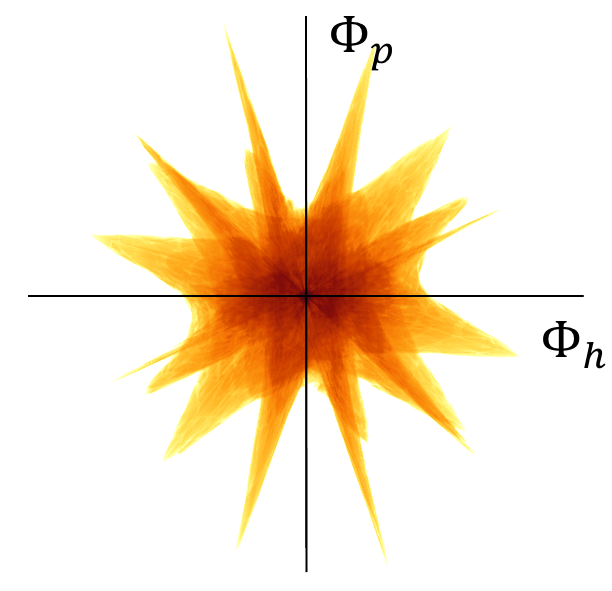}
    &
    \includegraphics[width=0.13\textwidth]{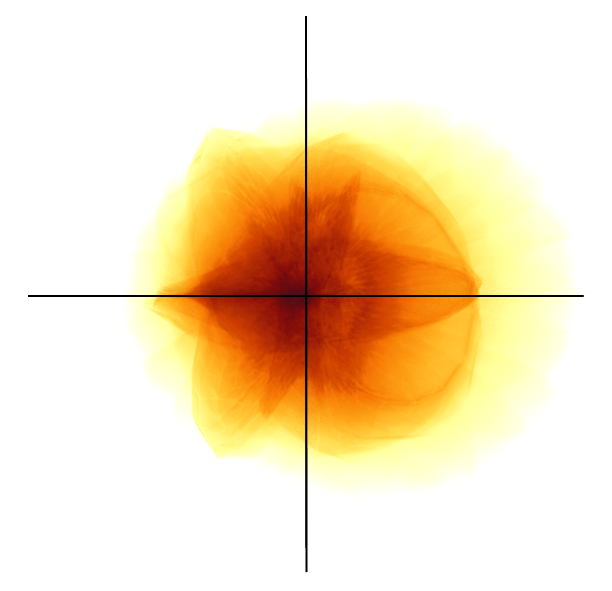}
    &
    \includegraphics[width=0.13\textwidth]{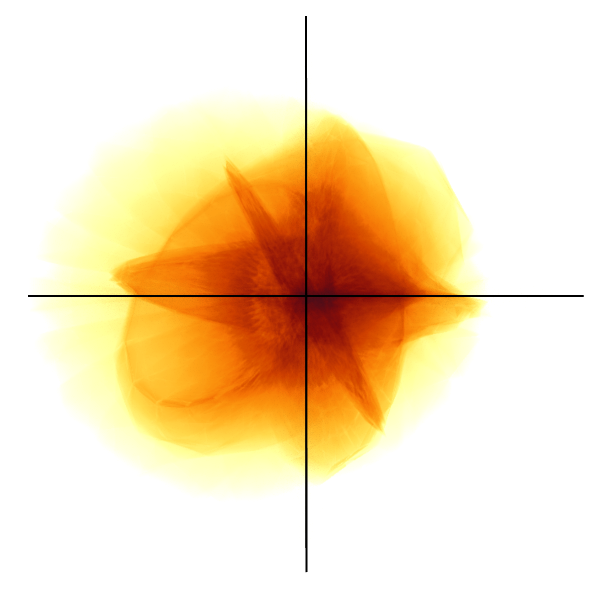}
    &
    \includegraphics[width=0.13\textwidth]{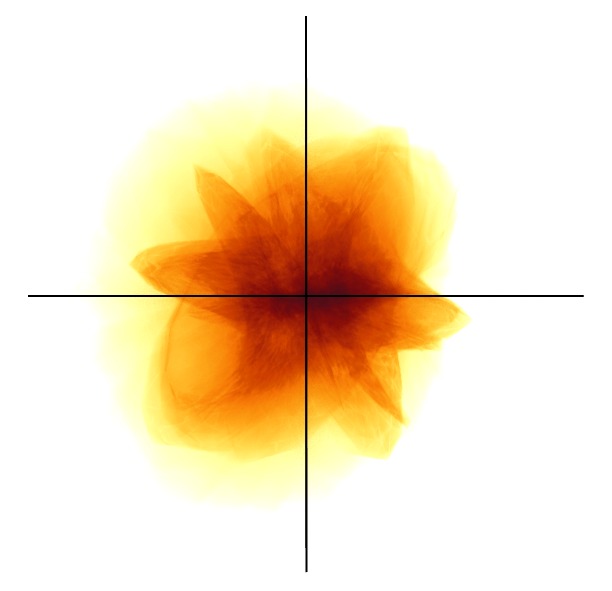}
    &
    \includegraphics[width=0.13\textwidth]{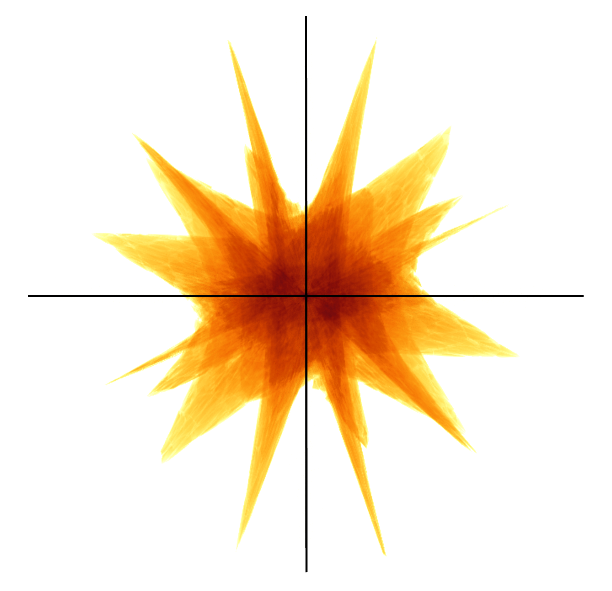}
    &
    \includegraphics[width=0.13\textwidth]{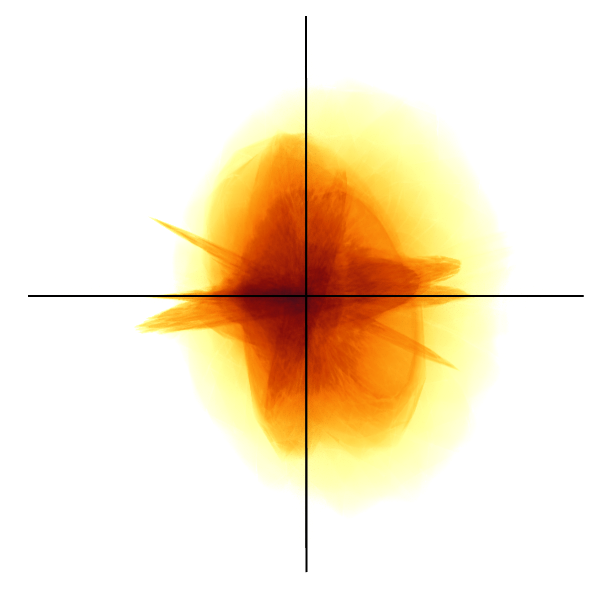}
    \\
    \raisebox{0.4\height}{\rotatebox{90}{Thiophene}}
    &
    \includegraphics[width=0.13\textwidth]{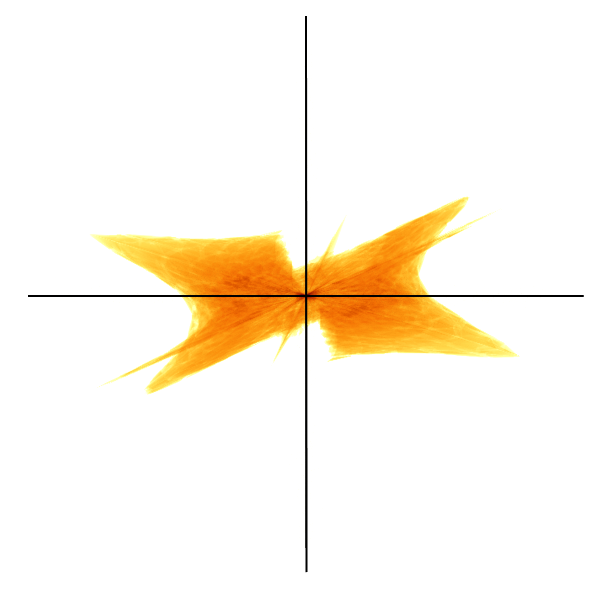}
    &
    \includegraphics[width=0.13\textwidth]{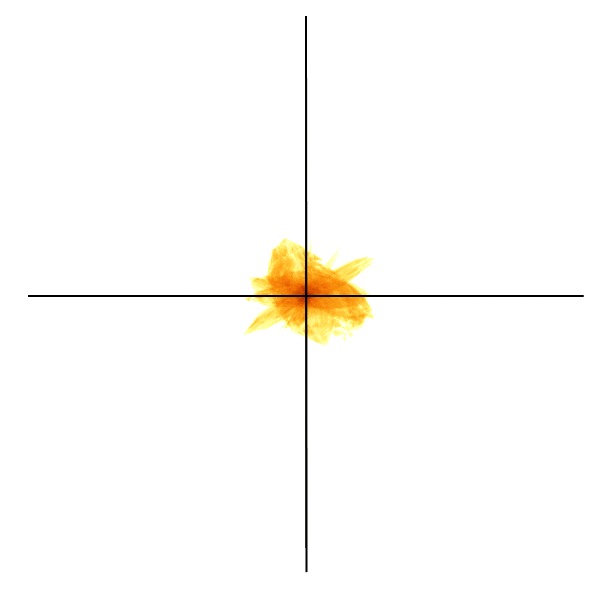}
    &
    \includegraphics[width=0.13\textwidth]{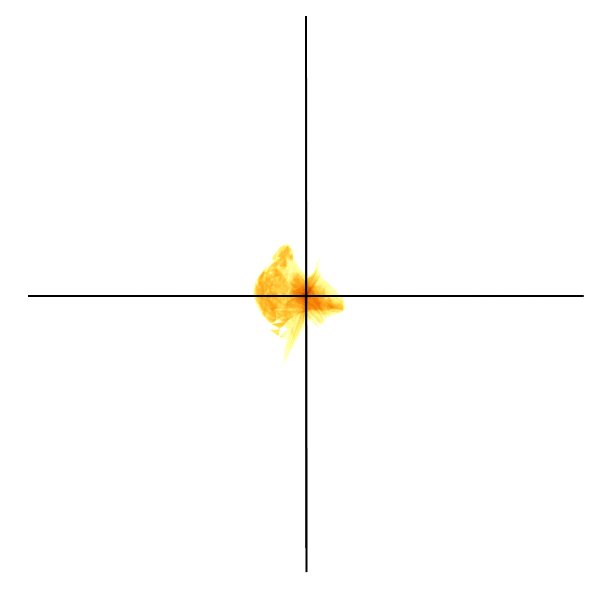}
    &
    \includegraphics[width=0.13\textwidth]{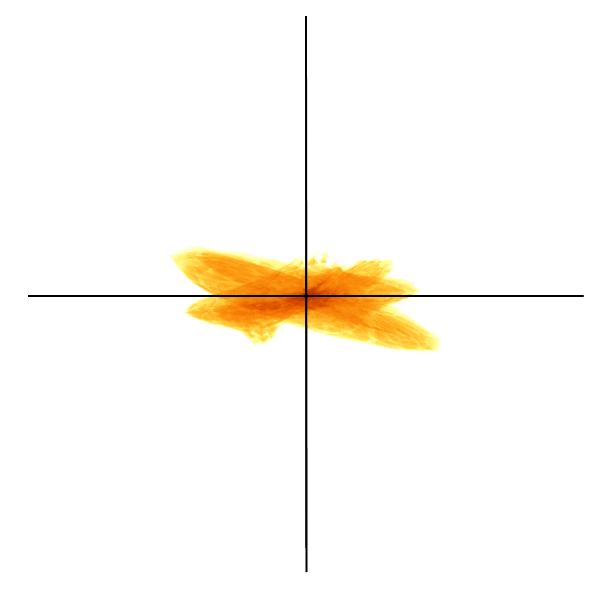}
    &
    \includegraphics[width=0.13\textwidth]{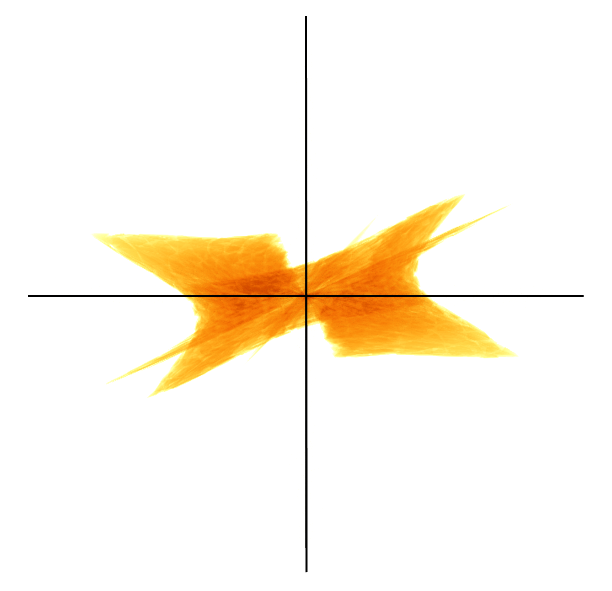}
    &
    \includegraphics[width=0.13\textwidth]{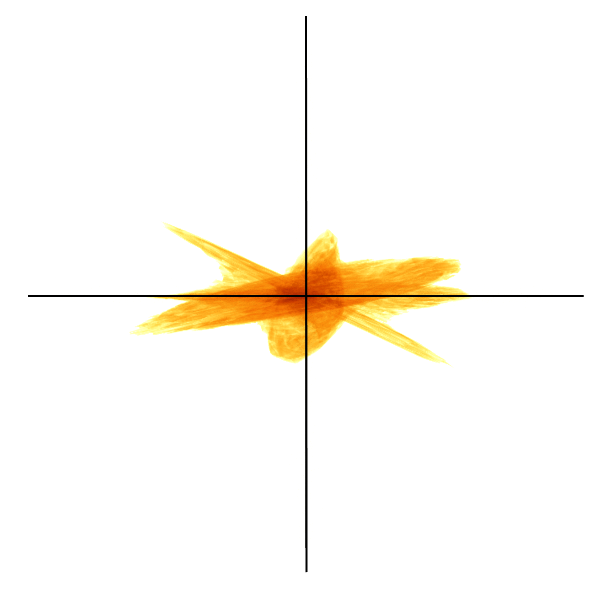}
    \\
    \raisebox{0.2\height}{\rotatebox{90}{Quinoxaline}}
    &
    \includegraphics[width=0.13\textwidth]{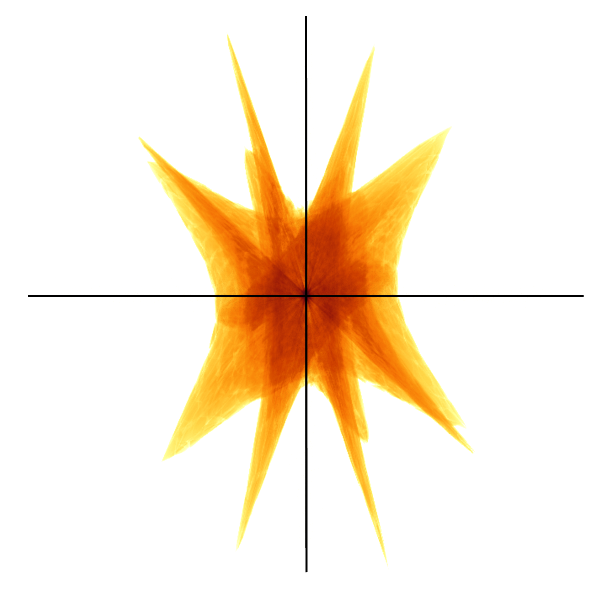}
    &
    \includegraphics[width=0.13\textwidth]{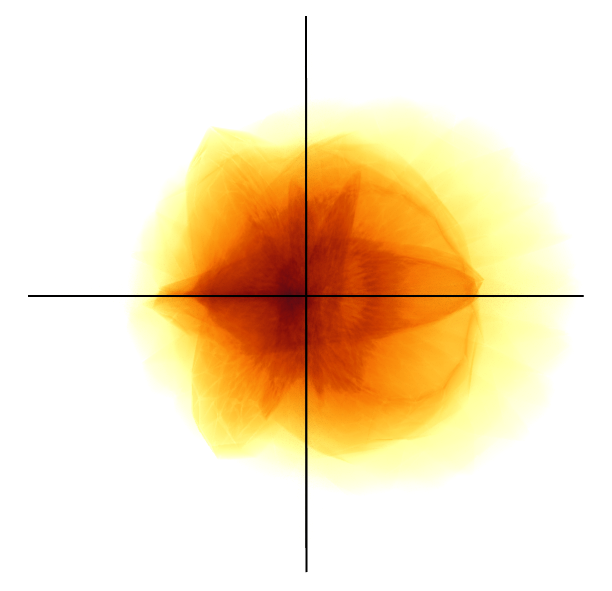}
    &
    \includegraphics[width=0.13\textwidth]{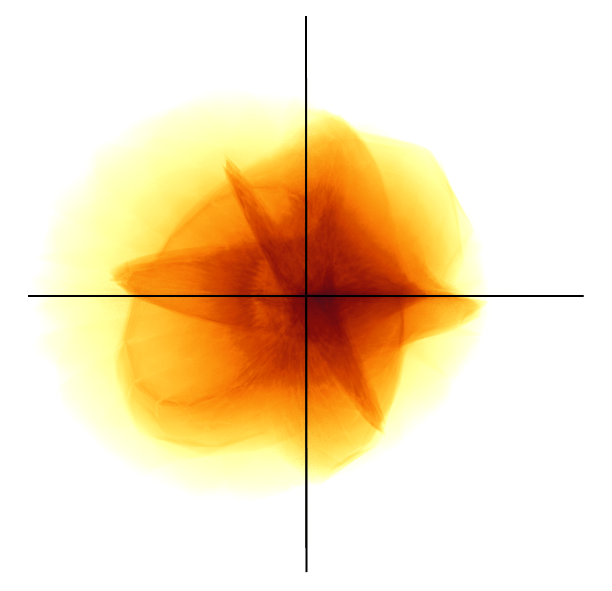}
    &
    \includegraphics[width=0.13\textwidth]{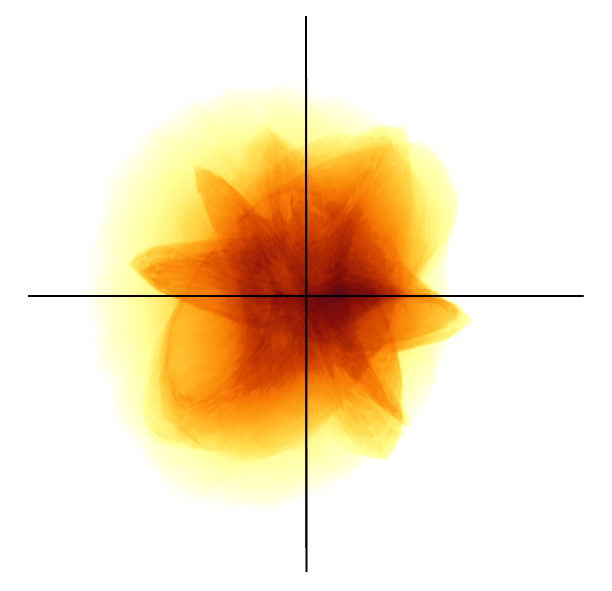}
    &
    \includegraphics[width=0.13\textwidth]{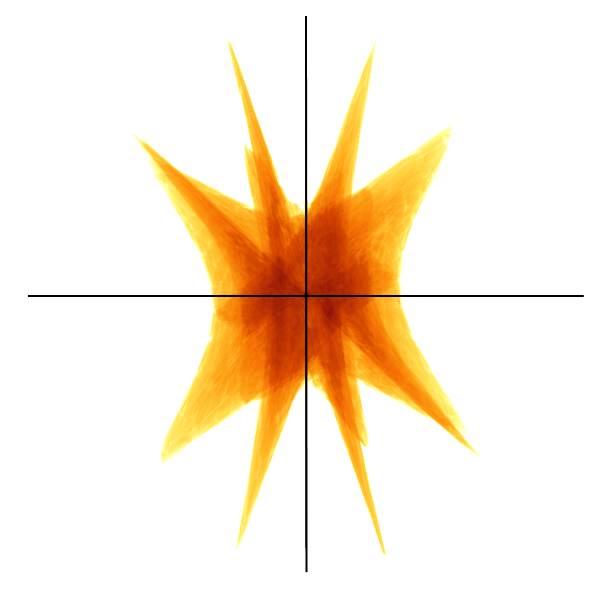}
    &
    \includegraphics[width=0.13\textwidth]{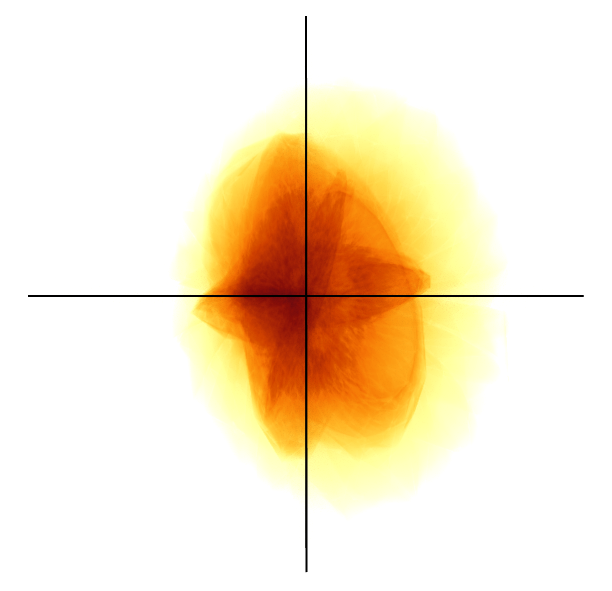}
    \end{tabular}
    \caption{CSP peeling for analyzing Thiophene-Quinoxaline (green and orange planes, respectively) for varying dihedral angle based conformations of state~1. In the $60^{\circ}$ and $90^{\circ}$ conformations, the peeled CSP for Thiophene covers a small region suggesting local excitation within Quinoxaline. $90^{\circ}$ exhibits least charge transfer from Thiophene to Quinoxaline. Thiophene behaves as a donor within other molecular conformations.} 
    \vspace{-1.2em}
    \label{fig:tq-geometry}
\end{figure*}

\begin{figure}[!ht]
    \centering
    \begin{tabular}{c@{\hskip3pt}c@{\hskip1pt}c@{\hskip1pt}c@{\hskip1pt}c@{\hskip1pt}}
    & \small{Cu-PHE-PHE} & \small{Cu-PHE-PHEOME} & \small{Cu-PHE-XANT}
    \\
    \cmidrule(lr){2-4}
    &
    \includegraphics[width=0.11\textwidth]{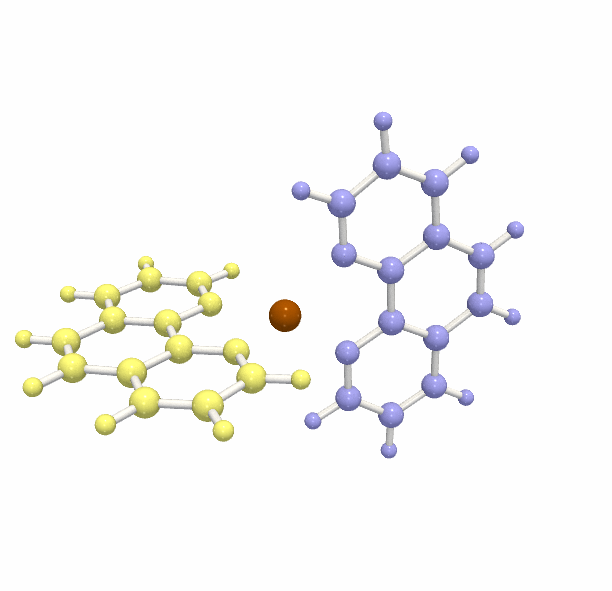}
    &
    \includegraphics[width=0.11\textwidth]{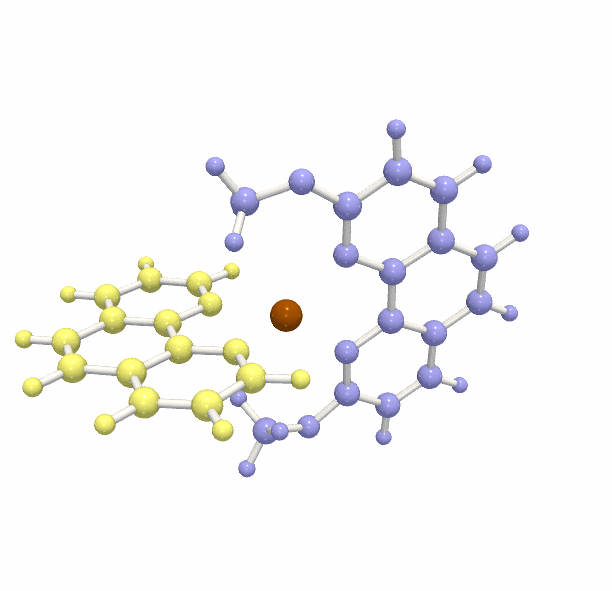}
    &
    \includegraphics[width=0.11\textwidth]{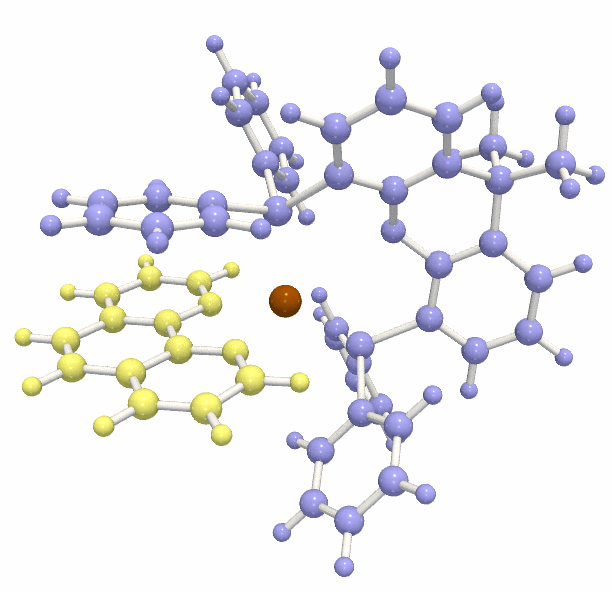}
    \\
    \raisebox{0.15\height}{\rotatebox{90}{\small{Complete CSP}}}
    &
    \includegraphics[width=0.11\textwidth]{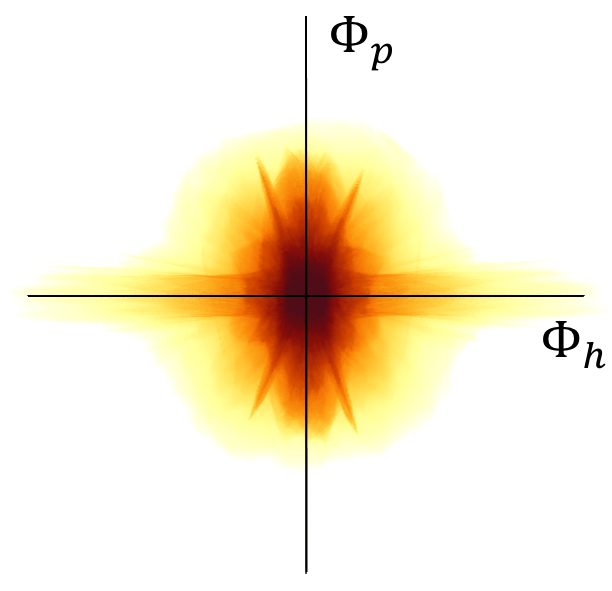}
    &
    \includegraphics[width=0.11\textwidth]{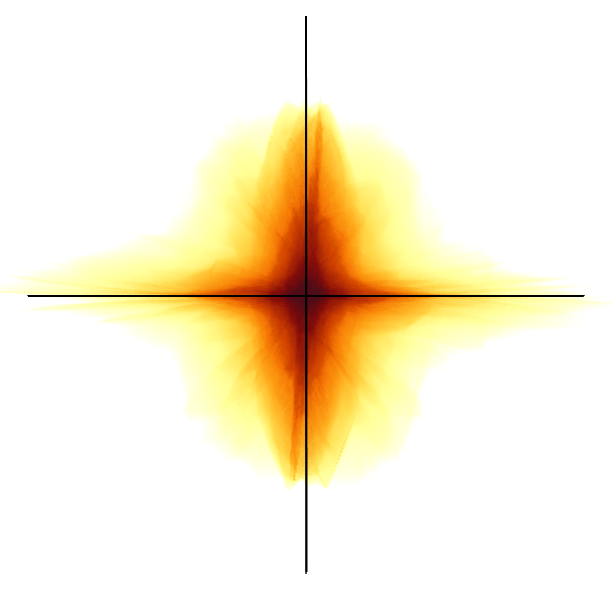}
    &
    \includegraphics[width=0.11\textwidth]{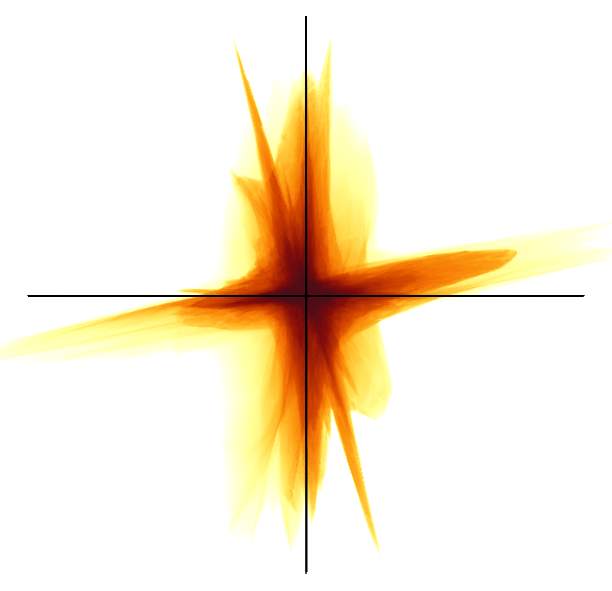}
    \\
    \raisebox{2.8\height}{\rotatebox{90}{\small{Cu}}}
    &
    \includegraphics[width=0.11\textwidth]{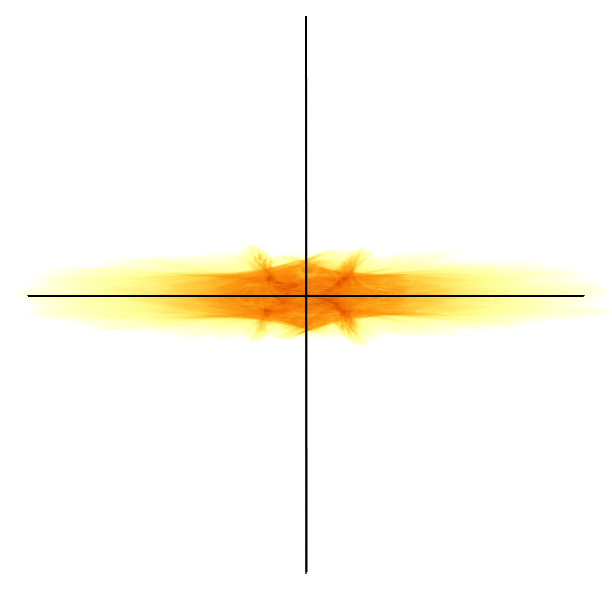}
    &
    \includegraphics[width=0.11\textwidth]{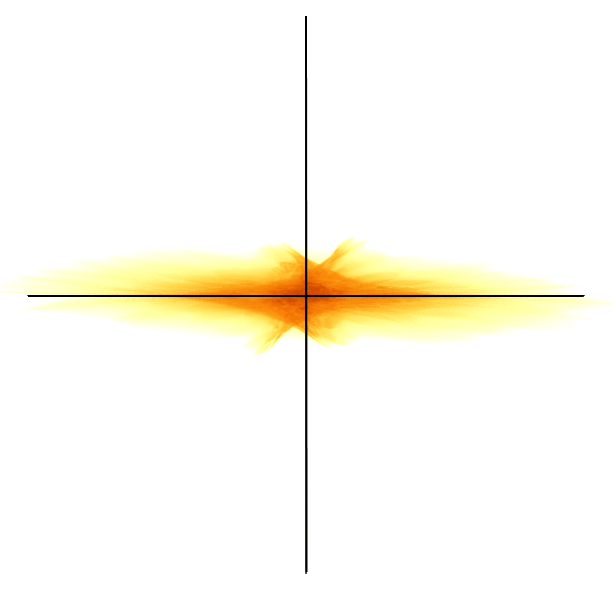}
    &
    \includegraphics[width=0.11\textwidth]{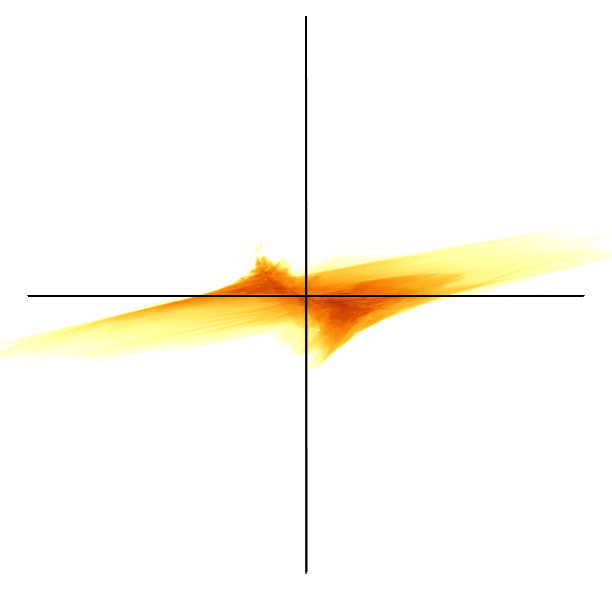}
    \\
    \raisebox{1.6\height}{\rotatebox{90}{\small{PHE}}}
    &
    \includegraphics[width=0.11\textwidth]{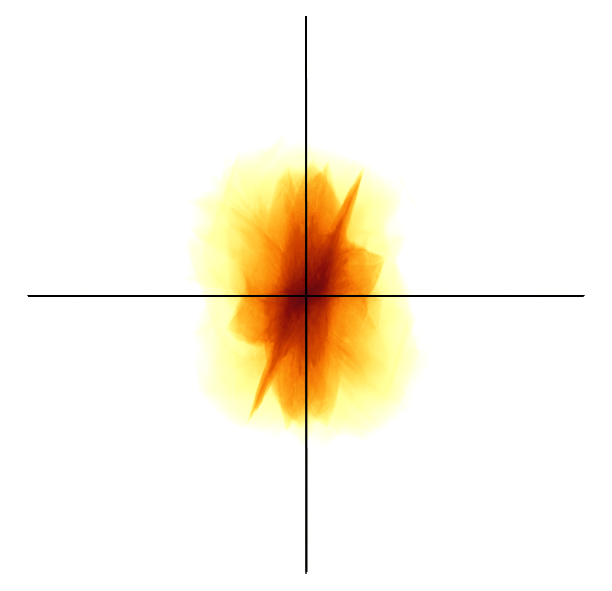}
    &
    \includegraphics[width=0.11\textwidth]{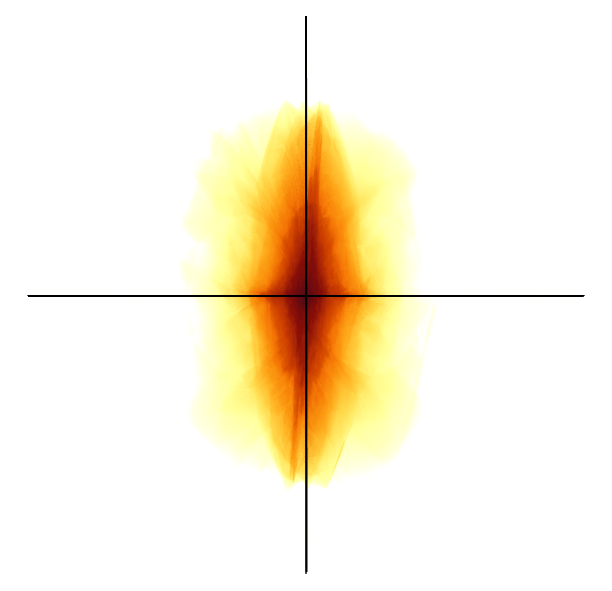}
    &
    \includegraphics[width=0.11\textwidth]{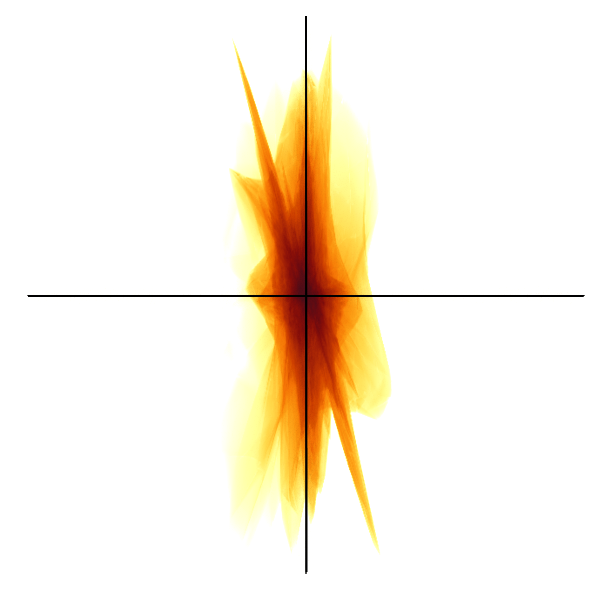}
    \\
    \raisebox{0.9\height}{\rotatebox{90}{\small{Ligand}}}
    &
    \includegraphics[width=0.11\textwidth]{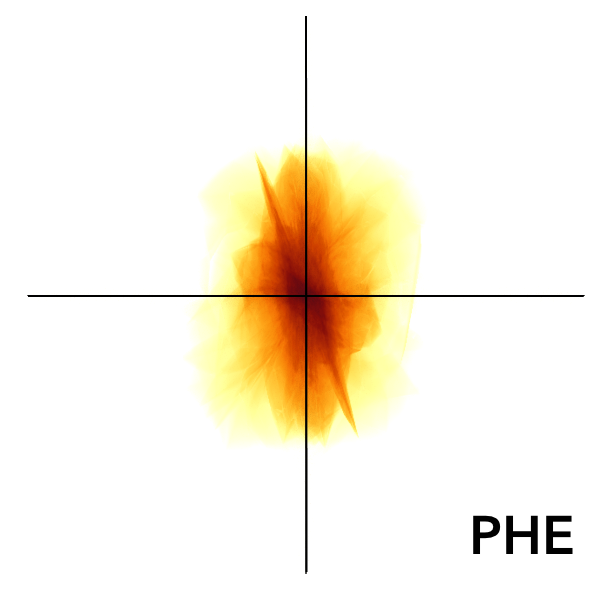}
    &
    \includegraphics[width=0.11\textwidth]{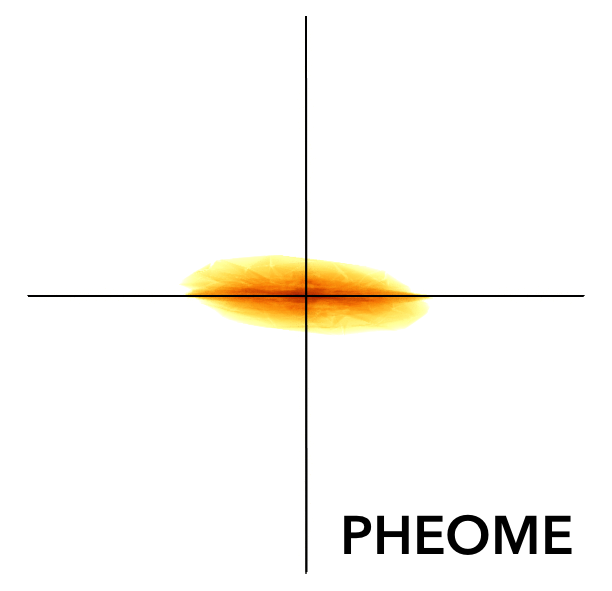}
    &
    \includegraphics[width=0.11\textwidth]{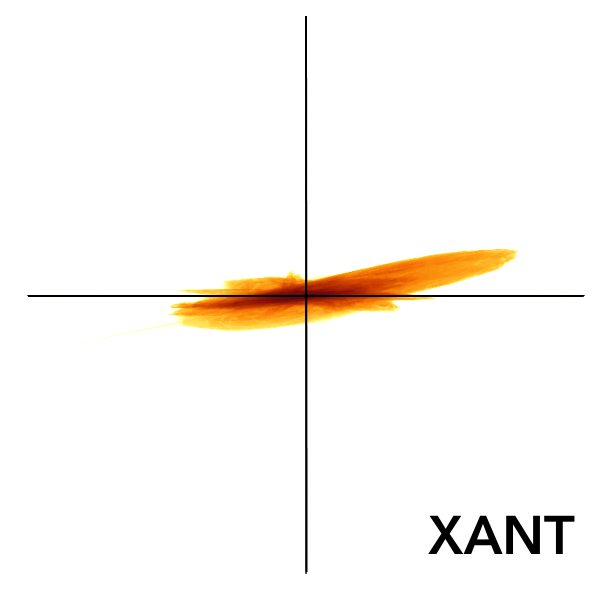}
   \end{tabular}
    \caption{CSP peeling for analyzing copper complexes with different ligand configurations, all in state~1. Copper behaves as a donor and PHE behaves as an acceptor within all complexes. The charge transfer from Cu is symmetric in Cu-PHE-PHE.} 
    \label{fig:cu-ligands}
\end{figure}

\begin{figure}[!ht]
    \centering
    \begin{tabular}{c@{\hskip3pt}c@{\hskip1pt}c@{\hskip1pt}c@{\hskip1pt}c@{\hskip1pt}cc@{\hskip1pt}c}
    & \multicolumn{2}{c}{\small{Cu-PHE-PHEOME}} &\multicolumn{2}{c}{\small{Cu-PHE-XANT}} \\
    \cmidrule(lr){2-3} \cmidrule(lr){4-5}
    & \small{State 9} & \small{State 10} & \small{State 3} & \small{State 10}
    \\
    \cmidrule(lr){2-3} \cmidrule(lr){4-5}
    \raisebox{0.2\height}{\rotatebox{90}{\small{Complete CSP}}}
    &
    \includegraphics[width=0.11\textwidth]{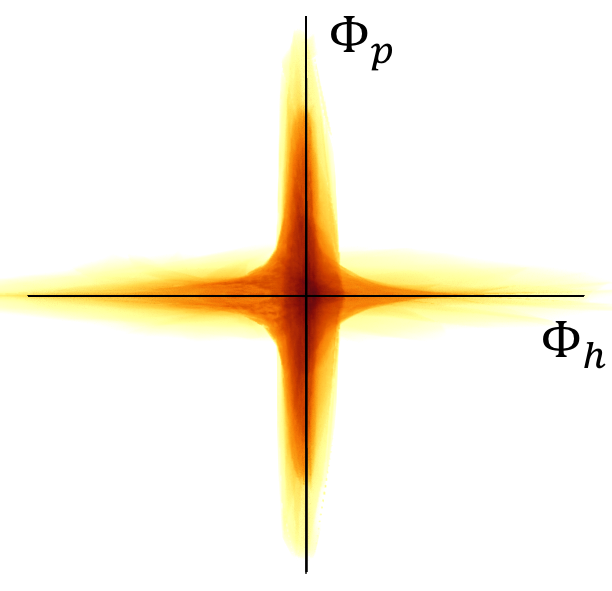}
    &
    \includegraphics[width=0.11\textwidth]{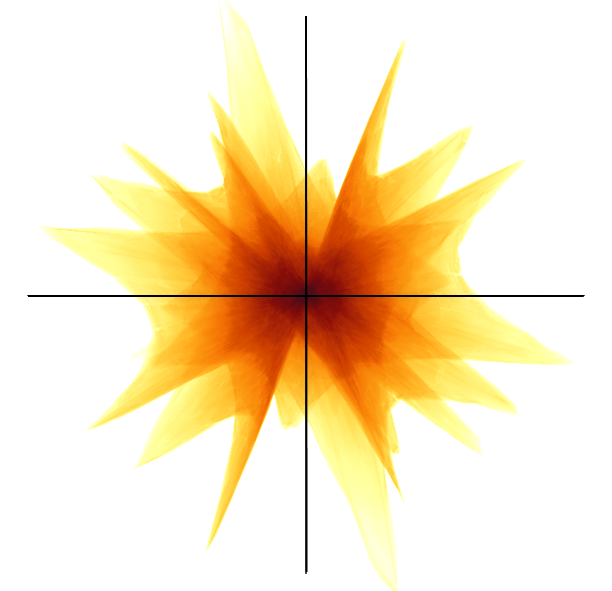}
    &
    \includegraphics[width=0.11\textwidth]{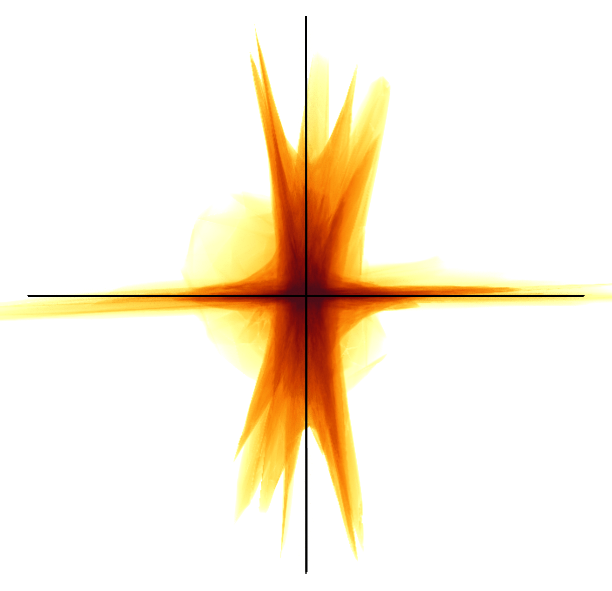}
    &
    \includegraphics[width=0.11\textwidth]{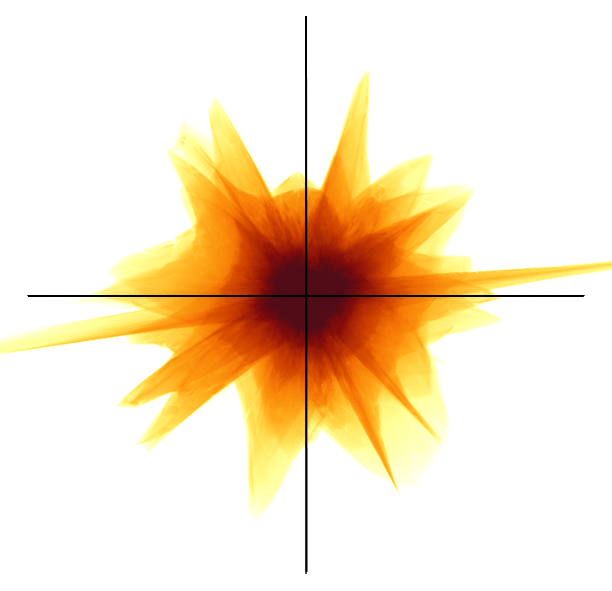}
    \\
    \raisebox{2.5\height}{\rotatebox{90}{\small{Cu}}}
    &
    \includegraphics[width=0.11\textwidth]{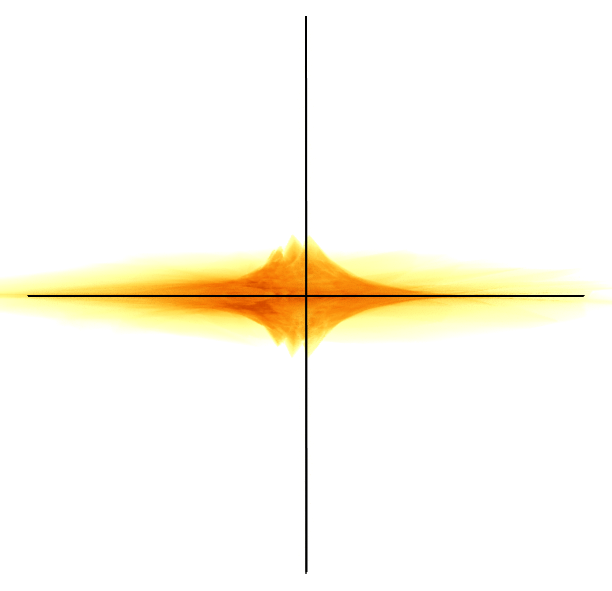}
    &
    \includegraphics[width=0.11\textwidth]{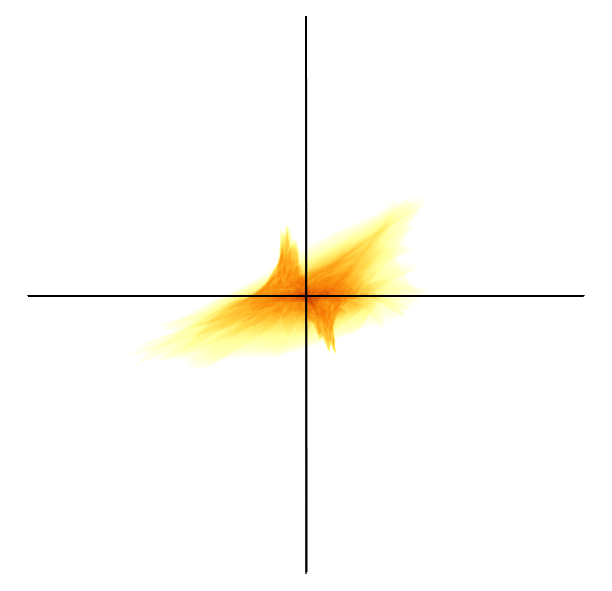}
    &
    \includegraphics[width=0.11\textwidth]{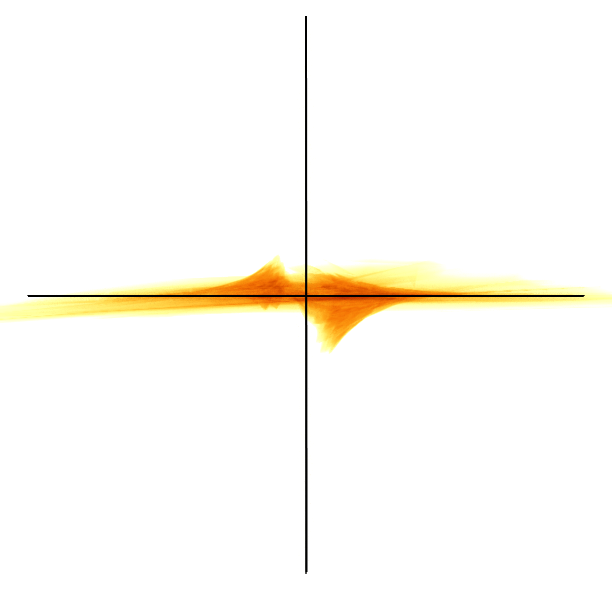}
     &
    \includegraphics[width=0.11\textwidth]{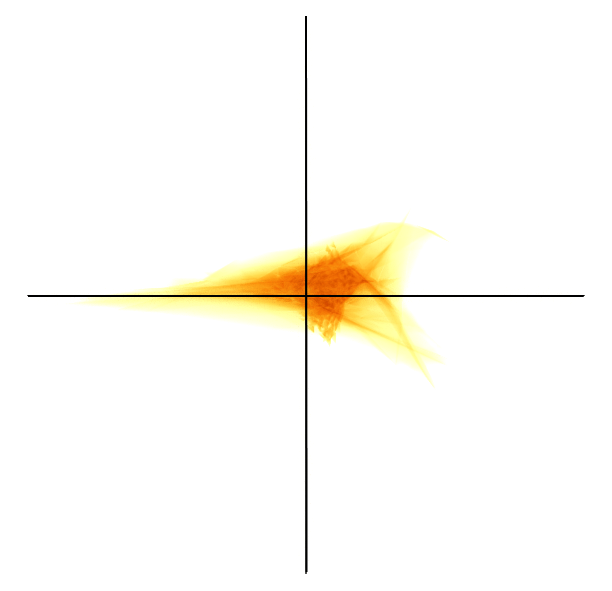}
    \\
    \raisebox{1.5\height}{\rotatebox{90}{\small{PHE}}}
    &
    \includegraphics[width=0.11\textwidth]{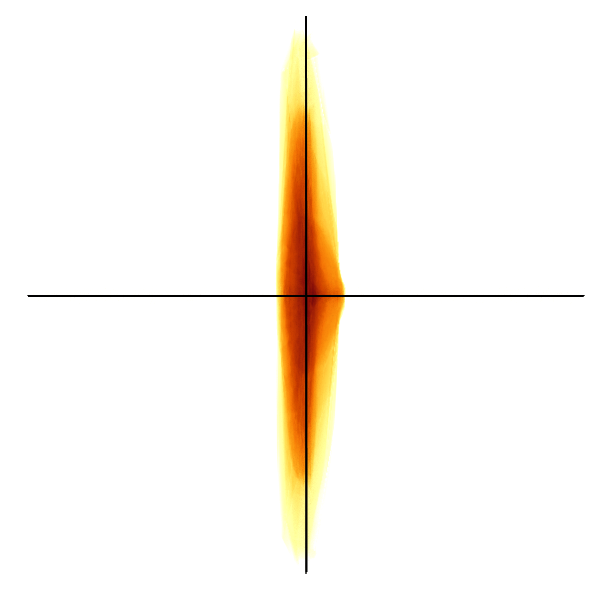}
    &
    \includegraphics[width=0.11\textwidth]{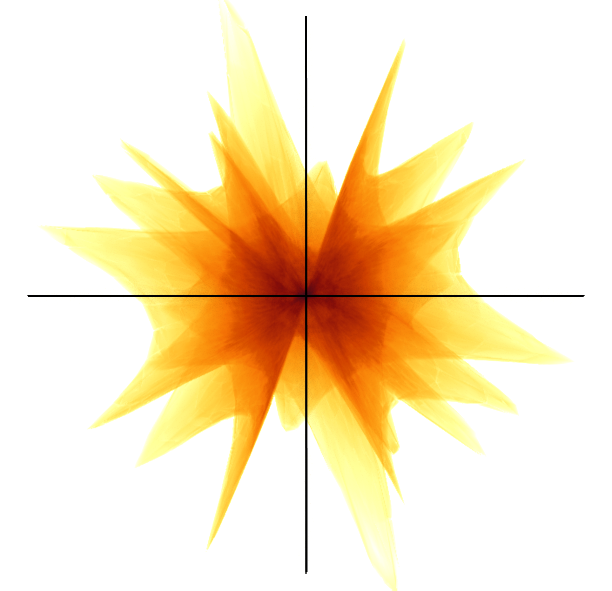}
    &
    \includegraphics[width=0.11\textwidth]{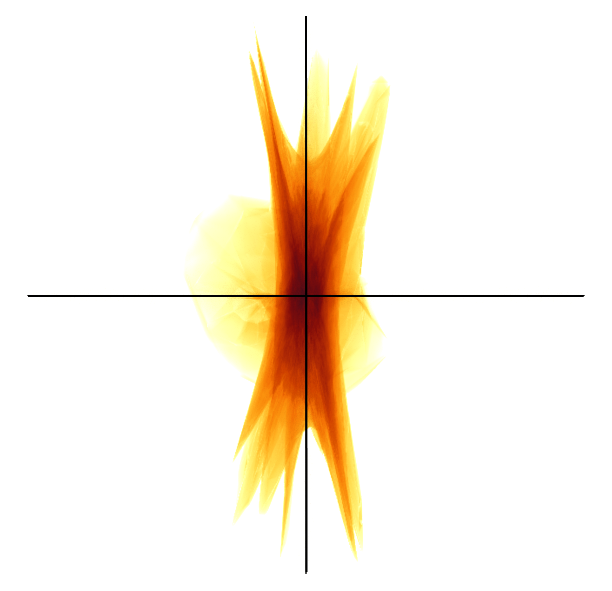}
    &
    \includegraphics[width=0.11\textwidth]{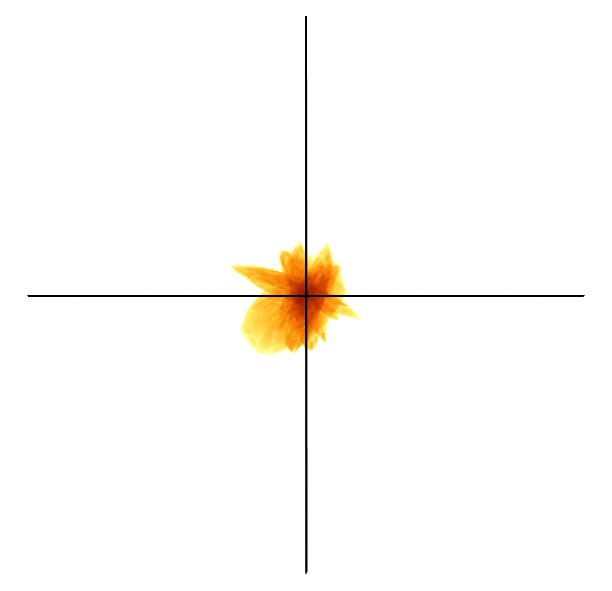}
    \\
    \raisebox{0.8\height}{\rotatebox{90}{\small{Ligand}}}
    &
    \includegraphics[width=0.11\textwidth]{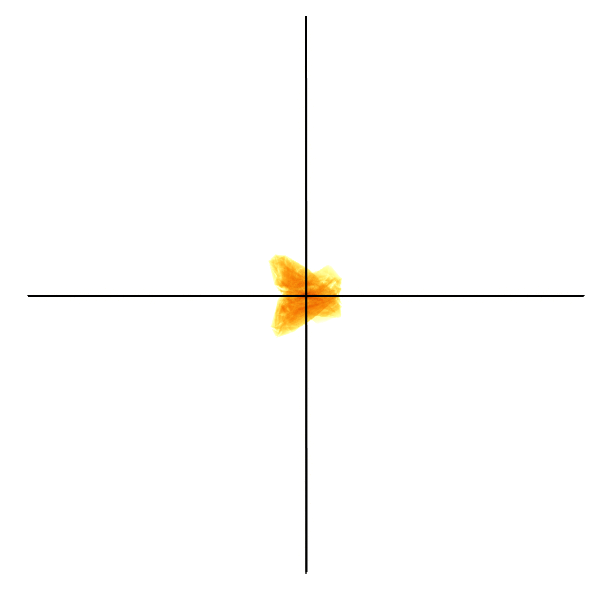}
    &
    \includegraphics[width=0.11\textwidth]{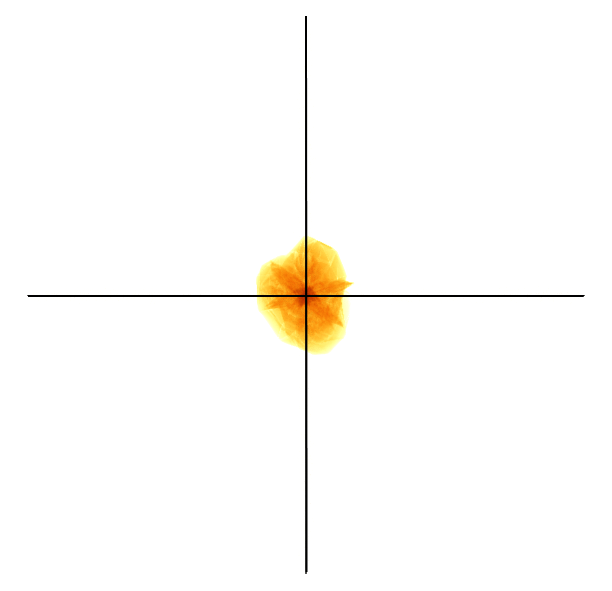}
    &
    \includegraphics[width=0.11\textwidth]{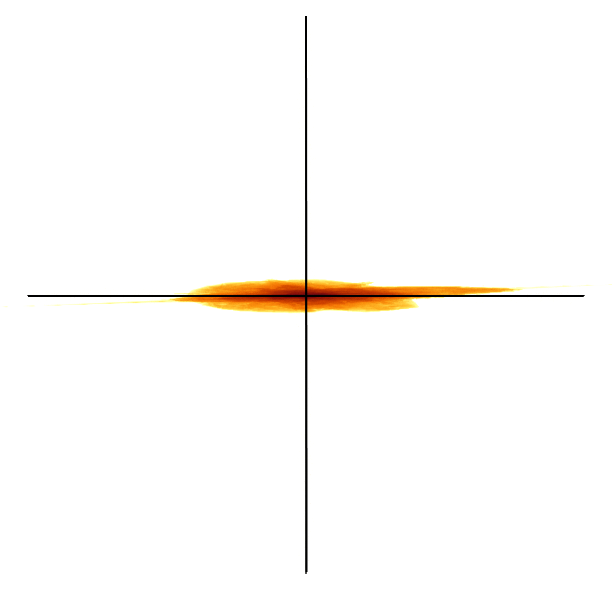}
    &
    \includegraphics[width=0.11\textwidth]{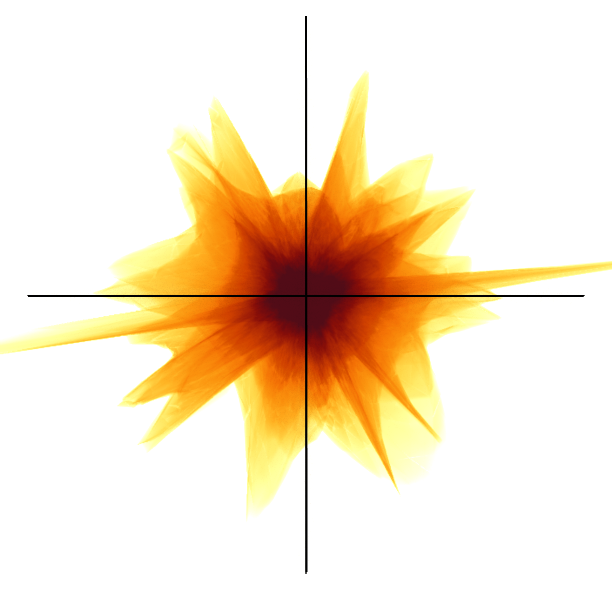}
   \end{tabular}
    \caption{CSP peeling for comparing different excitation states. For Cu-PHE-PHEOME, we observe charge transfer in state~9 and local excitation within PHE in state~10. For Cu-PHE-XANT, in state~3 we note charge transfer and local excitation within XANT in state~10.} 
    \label{fig:cu-lect}
    \vspace{-1em}
\end{figure}

\section{Bivariate analysis of electronic transitions}
\label{sec:bivariate}
Molecular orbitals provide an approximate solution for the electron distribution in molecules. Typically they are calculated using Density Functional Theory (DFT), applying programs like Gaussian~\cite{Frisch2016Gaussian}. DFT methods also support the simulation of electronic transitions within a molecule. 
The result is two scalar fields, the hole NTO ($\phi_h$) and  particle NTO ($\phi_p$), here treated as a bivariate field (x: $\phi_h$, y: $\phi_p$), over a 3D spatial domain for the molecule. 

Continuous scatter plots (CSP) together with fiber surfaces provide a valuable starting point for the exploration of such bivariate fields.
CSPs are density plots in range space generalizing the concept of scatterplots to spatially continuous input data. They are related to histograms and isosurface statistics~\cite{Bachthaler2008CSP}.
The density at a point $(x,y)$ in the range space, referred to as point density to distinguish it from the electron density field, is mapped to color. 
Fiber surfaces generalize isosurface to bivariate fields. They are composed of sets of fibers, which are preimages of points in the range of the bivariate field~\cite{carr2015fiber}. 
Control line segments in the CSP define the fiber surfaces in the spatial domain.
  
An example for the copper complex Cu-PHE-PHEOME, with its unique cross-shaped CSP is shown in \autoref{fig:teaser}. Two pairs of segments (red and green) across the spikes of the CSP have been selected, which results in fiber surfaces that cover regions across individual atoms of the molecule~\autoref{fig:teaser}~(b). 
The green fiber surface covers the PHE ligand and the red fiber surface the copper atom at the centre. 
These observations and further exploration of the data showed that an atom's behavior as donor or acceptor is encoded within the patterns of the CSP which acts as a descriptor for the electronic transitions. 
This motivated the idea of CSP peeling, which gives direct access to transition patterns of the individual atoms or atomic groups and eases the exploration.
It displays the contributions of selected subgroups to the CSP in comparison with each other. The example in \autoref{fig:teaser} is simple with dominant and clear spikes along the two axes and the fiber surfaces are similar to isosurfaces for corresponding isovalues of $\phi_h$ and $\phi_p$. However, the CSP shapes are not simple in general, and require the peeling operator and more complex fiber surfaces for its systematic exploration. We consistently use the same axes and color map (log scale) in all figures within this paper.

\section{Visual Analysis Pipeline}
\autoref{fig:pipeline} presents an overview of the visual analysis pipeline that processes the input electron density fields together with the molecular structure (atom locations, radii, bond information). 

\myparagraph{Segmentation.}
A molecular subgroup is represented by a subset of atoms that constitute the molecule. Following previous work~\cite{masood2021visual}, a weighted Voronoi tessellation~\cite{Aurenhammer1987powerdiag} of a point set is computed to partition the molecule into atomic regions. The point location and weights correspond to the atom location and radii. A molecular subgroup is represented as a union of atomic regions.  Other geometric or topology-based segmentation may be used if they are deemed appropriate.  Atoms that constitute the various subgroups of interest are assumed to be available as input to the visualization pipeline.

\myparagraph{Continuous scatter plot (CSP) and fiber surfaces.}
The CSP of the bivariate field ($\phi_h$, $\phi_p$) is computed and displayed to provide an overview. It may be explored interactively by specifying query line segments and extracting the corresponding fiber surfaces, the preimage of the line segments. Extraction and visualization of fiber surfaces is one approach towards studying and understanding the domain and range space of the bivariate field, \autoref{sec:bivariate}. 

\myparagraph{CSP peeling.}
A detailed analysis of the CSP is provided via CSP peeling. This is the computation of CSP for the field restricted to a region of interest within the spatial domain to study and compare characteristic patterns, \autoref{sec:interpretation}. 
    
\section{Interpreting a Peeled CSP}
\label{sec:interpretation}
If the peeled CSP aligns with the X-axis or specifically consists of the region satisfying $| \phi_h | > | \phi_p |$, then we classify the subgroup as a donor. A horizontal line $\phi_p = 0$ implies that the particle NTO equals zero within the corresponding region in the domain and there is no charge gain. Rather, this region has lost charge because hole NTO $\neq$ 0. Similarly, if the CSP aligns with the Y-axis, then the corresponding region in the domain represents an acceptor. In \autoref{fig:teaser}, Cu and PHEOME are classified as donors while PHE as an acceptor.

A large difference in the area covered by the CSP of subgroups may help infer which subgroup is a stronger donor or acceptor, as appropriate. Alternatively, a consistent CSP for a subgroup within different complexes indicates a unique property of the subgroup. For example, we compare the behavior of copper within different complexes in \autoref{fig:cu-ligands}.
Peeling also helps identify the nature of the electronic transition. Local excitation refers to the scenario where both hole and particle NTOs are located within a subgroup, and charge transfer excitation where they are located within different subgroups. 
Local excitation may be identified by comparing CSPs of individual subgroups. If the contribution from all subgroups except one is small then we may conclude that the electronic transition is local to a subgroup. For example, we identify State~10 as a local excitation state in Cu-PHE-PHEOME, see \autoref{fig:cu-lect}.

\section{Results}
We now present results of two case studies where CSP peeling reveals the donor-acceptor behavior, how it varies with geometric conformations or with varying ligands in a family of complexes. The hole and particle NTOs for these case studies are calculated using the Gaussian software package~\cite{Frisch2016Gaussian} while the CSPs and fiber surfaces are computed using the topology toolkit~\cite{Tierny2018ttk}.

\subsection{Case Study 1: Thiophene-Quinoxaline}
The first case study investigates different geometric conformations of a molecule that can be divided into two subgroups, Thiophene and Quinoxaline, first row in~\autoref{fig:tq-geometry}. Our analysis focuses on the first excited state of the molecule.
Thiophene (the 5-member ring) is generally considered to be a donor subgroup but, as our study shows, this property varies depending on the dihedral angle between the subgroups. The $\pi$-bond conjugation (delocalization of electrons over the molecule)  can be broken (at 90°) which results in different types of excitation, local vs. charge transfer, for different angles. The CSP analysis results are shown in \autoref{fig:tq-geometry}.

The $\pi$-bond conjugation is maximum in $0^{\circ}$ and $180^{\circ}$ as the orbitals are fully delocalized over the molecule. The CSPs for these two angles are near identical, and hence capture the similarity between the two conformations. The peeled CSPs corresponding to Thiophene for $0^{\circ}$ and $180^{\circ}$ are aligned along the X-axis implying that Thiophene is the donor subgroup, the Quinoxaline CSPs are more aligned along the Y-axis suggesting that it is the acceptor in both conformations.
The area covered by Thiophene CSP shrinks towards origin from $0^{\circ}$ to $60^{\circ}$ and reduces to a minimum at $90^{\circ}$. This suggests that both hole and particle NTOs for Thiophene reduce and hence the charge transfer from Thiophene to Quinoxaline also decreases as the dihedral angle increases. 
Such a behavior is expected because the $\pi$-bond conjugation is minimum at $90^{\circ}$. The area increases when the dihedral angle increases further to $120^{\circ}$.
Quinoxaline CSP for $60^{\circ}$ and $90^{\circ}$ is similar to the CSP of the molecule, while Thiophene CSP covers a small region near the origin. This suggests  local excitation in these conformations, primarily within Quinoxaline. The behavior changes towards charge transfer excitation again at $120^{\circ}$.
In the relaxed geometry, where the molecule is at an energy optimal dihedral angle ($35.3^{\circ}$), we observe that Thiophene acts as a donor, which is again the expected behavior.

\subsection{Case Study 2: Copper complexes}
In this case study, we consider copper complexes consisting of a copper subgroup, a fixed ligand (PHE), and a second ligand that varies between PHE, PHEOME and XANT. 

\myparagraph{Varying ligands.}
First, we study the effect of varying the second ligand with a focus on the first excited state of each molecule, see \autoref{fig:cu-ligands}.
Copper behaves as a donor in all configurations with similar strength. This is an expected behavior for copper.
PHE behaves as an acceptor. An interesting observation is that the charge transfer is expected to be symmetric in Cu-PHE-PHE, from Cu to both PHE ligands. This behavior is observable from the CSPs of the individual PHE ligands.
In Cu-PHE-PHEOME, both Cu and PHEOME are donors. Cu is expected to be a stronger donor. Copper covers a larger span on the X-axis indicating larger values of $\phi_h$ in its atomic region. However, a visual inspection of the CSPs is not sufficient to claim that copper is indeed the stronger donor. CSPs for Cu and PHEOME appear to be similarly close to origin, but a quantification step is necessary in order to confirm that the point densities are indeed similar. 
In Cu-PHE-XANT, both Cu and XANT behave as donors. XANT is expected to be a stronger donor than Cu but a visual inspection is again not sufficient to make such a claim.

\myparagraph{Characterizing excitations.}
We now compare the behavior of two complexes, Cu-PHE-PHEOME and Cu-PHE-XANT, within different excitation states. The aim is to characterize the nature of excitation, local or charge transfer. \autoref{fig:cu-lect} shows the results of the CSP peeling based visual analysis. 
For Cu-PHE-PHEOME, we observe a charge transfer excitation in state~9. The Cu and PHE CSPs exhibit donor and acceptor behaviors, respectively. We may conclude that a significant charge transfer happens from Cu to PHE. However, in state~10, CSPs of the molecule and that of PHE are similar. Further, contributions from other subgroups appear to be low indicating local excitation within PHE. In state~3 of Cu-PHE-XANT, we observe a charge transfer from Cu and XANT to PHE, and in state~10 a local excitation within XANT.

\section{Conclusions}
This paper presented a simple and effective approach to bivariate data analysis for visually interpreting electronic transition data. The case studies demonstrated the use of CSP peeling for identifying and comparing donor-acceptor behavior of molecular subgroups across different geometric conformations. Topics for future work include investigation of other segmentation methods and quantification of the peeled CSPs.


\acknowledgments{
\small{This work is partially supported by an Indo-Swedish joint network project: DST/INT/SWD/VR/P-02/2019 and VR grant 2018-07085, MHRD Govt. of India, a Swarnajayanti Fellowship from DST India (DST/SJF/ETA-02/2015-16), a Mindtree Chair research grant, the SeRC (Swedish e-Science Research Center), the Swedish Research Council~(VR) grant 2019-05487. The computations were enabled by resources provided by the Swedish National Infrastructure for Computing (SNIC) at NSC partially funded by the VR grant agreement no. 2018-05973.}}

\bibliographystyle{abbrv-doi-hyperref}

\bibliography{references}
\end{document}